# An Enhancement of Decimation Process using Fast Cascaded Integrator Comb (CIC) Filter

Rozita Teymourzadeh, IEEE Student Member, Masuri Bin Othman
VLSI Design Center, Institute of Microengineering and Nanoelectronics (IMEN) Universiti Kebengsaan Malaysia, 43600 Bangi, Selangor, Malaysia
rozita60@vlsi.eng.ukm.my

*Abstract -* **The over sampling technique has been shown to increase the SNR and is used in many high performance system such as in the ADC for audio and DAT systems. This paper presents the design of the decimation and its VLSI implementation which is the sub-component in the over sampling technique. The design of three main units in the decimation stage that is the Cascaded Integrator Comb (CIC) filter, the associated half band filters and the droop correction are also described. The Verilog HDL code in Xilinx ISE environment has been derived to describe the CIC filter properties and downloaded in to Virtex II FPGA board. In the design of these units, we focus on the trade-off between the speed improvement and the power consumption as well as the silicon area for the chip implementation.**

## I. INTRODUCTION

The most popular A/D converters for audio applications are realized based on the use of over sampling and sigma-delta ($\sum\Delta$) modulation techniques followed by decimation process [1]. Oversampled Sigma delta ($\sum\Delta$) modulator provides high resolution sample output in contrast to the standard Nyquist sampling technique. However at the output, the sampling process is needed in order to bring down the high sampling frequency and obtain high resolution. The CIC filter is a preferred technique for this purpose. In 1981, Eugene Hogenauer [2] invented a new class of economical digital filter for decimation called a Cascaded Integrator Comb filter (CIC) or recursive comb filter. This filter worked with sampling frequency of 5 MHz. Additionally the CIC filter does not require storage for filter coefficients and multipliers as all coefficients are unity [3]. Furthermore its on-chip implementation is efficient because of its regular structure consisting of three basic building blocks, minimum external control and less complicated local timing is required and its change factors is reconfigurable with the addition of a scaling circuit and minimal changes to the filter timing. It is also used to perform filtering of the out of band quantization noise and prevent excess aliasing introduced during sampling rate decreasing. Hence enhanced high speed will be key issue in chip implementation of CIC decimators. In 1998, Garcia [4] designed Residue Number System (RNS) for pipelined Hogenauer CIC. Compared to the two's complement design, the RNS based Hogenaur filter enjoys an improved speed advantage by approximately 54%. Similar structure by Meyer-Baese [5] has been implemented to reduce the cost in the Hogenauer CIC filter which shows that the filter can operate up to maximum clock frequency of 164.1 MHz on Altera FPLD and 82.64 MHz on Synopsys cell-based IC design.

This paper shows the implementation of the high speed CIC filters which are consist of three parts, integrator, comb and down sampler. The CIC filter is considered as recursive filter because of the feedback loop in integrator circuit and it can work with maximum throughput of 190 MHz.

The next section describes the mathematical formulation and block diagram of CIC filters in detail. Enhanced high speed architecture is explained in section III. Section IV shows implementation and design result in brief. Finally conclusion is expressed in section V.

## II. DEVELOPMENT OF A DECIMATION FILTER

The purpose of the CIC filter is twofold; firstly to remove filtering noise which could be aliased back to the base band signals and secondly to convert high sample rate m-bit data stream at the output of the Sigma-delta modulator to n-bit data stream with lower sample rate. This process is also known as decimation which is essentially performing the averaging and a rate reduction functions simultaneously.

Figure 1 shows the decimation process using CIC filter.

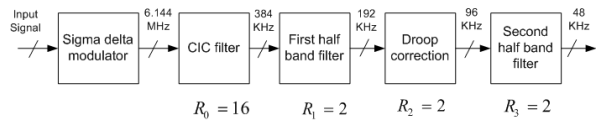

Fig. 1 Digital Decimation Process

The two half band filters [6] are used to reduce remain sampling rate reduction to the Nyquist output rate. First half band filter and second half band filter make the frequency response more flat and sharp similar to ideal filter frequency response.

Droop correction filter is allocated to compensate pass band attenuation which is created by the CIC filter. The frequency response of overall system will be shown in section V.

Table 1 shows filter specification in decimation process.

TABLE I
FILTER SPECIFICATIONS

| | Pass band (kHz) | Stop band (kHz) | Transition band (kHz) |
|---|---|---|---|
| CIC filter | 7 | 384 | 377 |
| First half band filter | 32 | 170 | 138 |
| Droop Correction | 32 | 70 | 38 |
| Second half band filter | 21.77 | 26.53 | 4.76 |

### III. PRINCIPLE OF CIC FILTER STRUCTURE

The CIC filter consist of N stages of integrator and comb filter which are connected by a down sampler stage as shown in figure 1 in z domain. The CIC filter has the following transfer function:

$$H(z) = H_I^N(z).H_C^N(z) = \frac{(1-z^{-RM})^N}{(1-z^{-1})^N} = \left( \sum_{k=0}^{RM-1} z^{-k} \right)^N \quad (1)$$

where N is the number of stage, M is the differential delay and R is the decimation factor.

In this paper, N, M and R have been chosen to be 5, 1 and 16 respectively to avoid overflow in each stages.

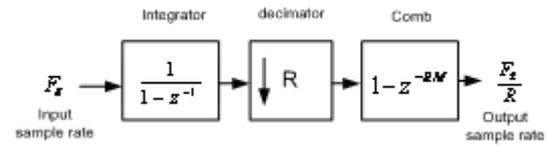

Fig. 2 One-stage of CIC filter block diagram

N, M and R are parameters to determine the register length requirements necessary to assure no data loss. Equation (1) can be express as follow:

$$H(z) = \sum_{k=0}^{(RM-1)N} h(k)z^{-k} = \left[ \sum_{k=0}^{RM-1} z^{-k} \right]^N \leq \left| \sum_{k=0}^{RM-1} z^{-k} \right|^'$$

$$\leq \left( \sum_{k=0}^{RM-1} |z|^{-k} \right)^N = \left( \sum_{k=0}^{RM-1} 1 \right)^N = (RM)^N \quad (2)$$

From the equation, the maximum register growth/width, $G_{max}$ can be expressed as:

$$G_{max} = (RM)^N \quad (3)$$

In other word, $G_{max}$ is the maximum register growth and a function of the maximum output magnitude due to the worst possible input conditions [2].

If the input data word length is $B_{in}$, most significant bit (MSB) at the filter output, $B_{max}$ is given by:

$$B_{max} = [N \log_2 R + B_{in} - 1] \quad (4)$$

In order to reduce the data loss, normally the first stage of the CIC filter has maximum number of bit compared to the other stages. Since the integrator stage works at the highest oversampling rate with a large internal word length, decimation ratio and filter order increase which result in more power consumption and speed limitation.

### III. SPEED IMPROVEMENT

#### A. Truncation for low power & high speed

Truncation means estimating and removing Least Significant Bit (LSB) to reduce the area requirements on chip and power consumption and also increase speed of calculation. Although this estimation and removing introduces additional error, the error can be made small enough to be acceptable for DSP applications.

Figure 3 illustrates five stages of the CIC filter when $B_{max}$ is 25 bit so truncation is applied to reduce register width. Matlab software helps to find word length in integrator and comb section.

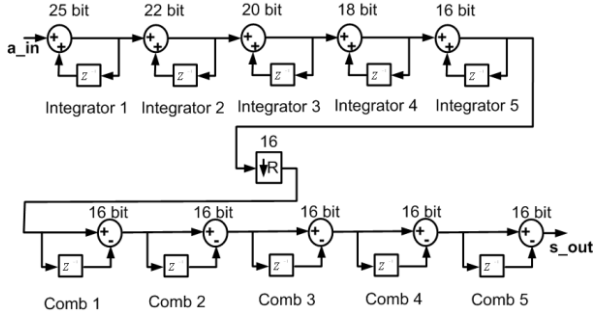

Fig. 3 Five-stages of truncated CIC filter

### B. Pipeline structure

One way to have high speed CIC filter is by implementing the pipeline filter structure. Figure 4 shows pipeline CIC filter structure when truncation is also applied. In the pipelined structure, no additional pipeline registers are used in integrator part. So that hardware requirement is the same as in the non-pipeline [7]. The CIC decimation filter clock rate is determined by the first integrator stage that causes more propagation delay than any other stage due to maximum number of bit. So it is possible to use a higher clock rate for a CIC decimation filter if a pipeline structure is used in the integrator stages, as compared to non-pipelined integrator stages. The clock rate in integrator section is R times higher than in the comb section.

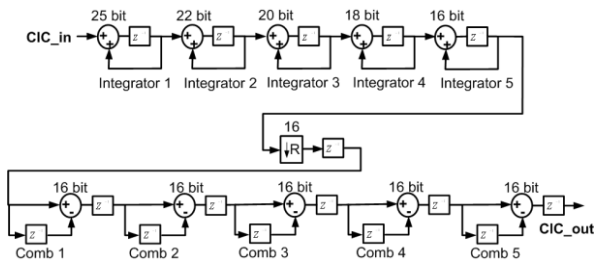

Fig. 4 Five-stage of truncated pipeline CIC filter

Previously, the pipeline structure for CIC filter was applied just for integrator part since the maximum clock rate is determined by the integrator. The above architecture showed that the maximum throughput was increased by 20 MHz when the pipeline structure is used for all the CIC parts consisting of integrator, comb and down sampler.

### C. Modified Carry look-ahead Adder (MCLA)

The other technique to increase speed is using Modified Carry Look-ahead Adder. The Carry Look-ahead adder (CLA) is the fastest adder which can be used for speeding up purpose but the disadvantage of the CLA adder is that the carry logic is getting quite complicated for more than 4 bits so Modified Carry Look-ahead Adder (MCLA) is introduced to replace as adder. This improve in speed is due to the carry calculation in MCLA. In the ripple carry adder, most significant bit addition has to wait for the carry to ripple through from the least significant bit addition. Therefore the carry of MCLA adder has become a focus of study in speeding up the adder circuits [8]. The 8 bit MCLA structure is shown in Figure 5. Its block diagram consists of 2, 4-bit module which is connected and each previous 4 bit calculates carry out for the next carry. The CIC filter in this paper has five MCLA in integrator parts. The maximum number of bit is 25 and it is decreased in next stages. So it truncated respectively to 25, 22, 20, 18 and 16 bit, left to right Notice that each 4-bit adder provides a group propagate and generate Signal, which is used by the MCLA Logic block. The group Propagate $P_G$ and Generate $G_G$ of a 4-bit adder will have the following expressions:

$$P_G = p_3.p_2.p_1.p_0 \qquad (5)$$

$$G_G = g_3 + p_3.g_2 + p_3.p_2.g_1 + p_3.p_2.p_1.g_0 \qquad (6)$$

The most important equations to obtain carry of each stage have been defined as below:

$$c_1 = g_0 + (p_0.c_0) \qquad (7)$$

$$c_2 = g_1 + (p_1.g_0) + (p_1.p_0.c_0) \qquad (8)$$

$$c_3 = g_2 + (p_2.g_1) + (p_2.p_1.g_0) + (p_2.p_1.p_0c_0) \qquad (9)$$

$$c_4 = g_3 + (p_3.g_2) + (p_3.p_2.g_1) + (p_3.p_2.p_1.g_0) + (p_3.p_2.p_1.p_0.c_0) \qquad (10)$$

Calculation of MCLA is based on above equations. 8-Bit MCLA Adder could be constructed continuing along in the same logic pattern, with the MSB carry-out resulting from OR & AND gates. The Verilog code has been written to implement addition. The MCLA Verilog code was downloaded to the Xilinx

FPGA chip. From Xilinx ISE synthesize report, it was found minimum clock period is 3.701ns (Maximum Frequency is 270 MHz).

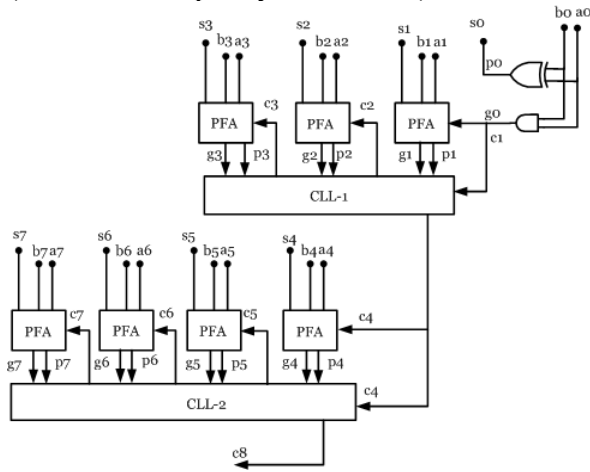

Fig. 5 The 8 bit MCLA structure

## V. IMPLEMENTATION

After the sigma delta modulator, the sampling rate must be reduced to 48 KHz which is the Nyquist sampling rate. This is carried out in 4-stages. The first stage involves the reduction of the sampling frequency by the decimation factor of 16. This is done by the CIC filter.

The remaining 3 stages involve the reduction of the sampling frequency by the decimation factor of 2 only which are carried out by the first half band, droop correction and the second half band respectively. Figure 6 illustrate the frequency response of the overall decimation filter when the sampling frequency is 6.144 MHz.

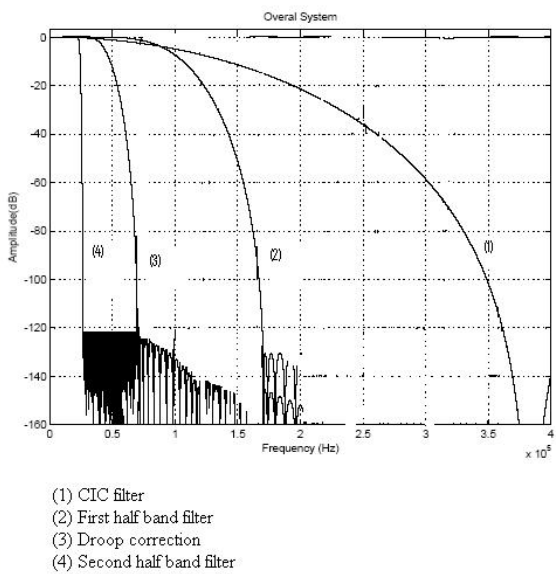

(1) CIC filter
(2) First half band filter
(3) Droop correction
(4) Second half band filter

Fig. 6 The frequency response of overall decimation filter

Figure 7 shows the Droop correction filter result. This filter design a low pass filter with pass band having the shape of inverse the CIC filter frequency response. So it compensates amplitude droop cause of the CIC filter and makes whole system frequency response flat.

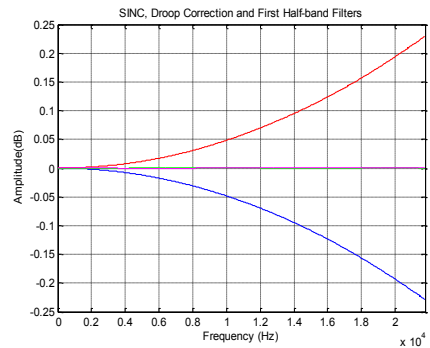

Fig. 7: Droop Correction effect on frequency response

Figure 8 shows the measured baseband output spectra before (Figure 8(a)) and after (Figure 8(b)) the decimation functions.

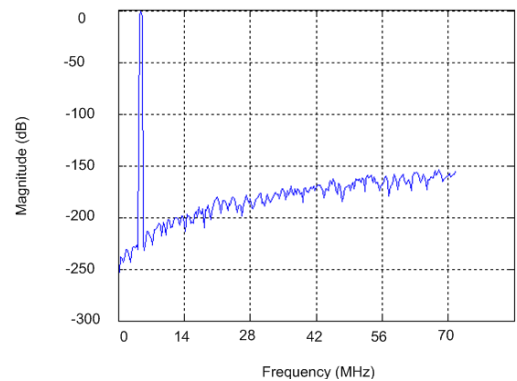

(a)

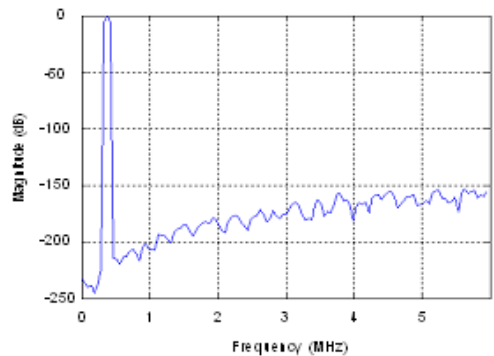

(b)

Fig. 8 Signal spectra (a) Output sigma delta modulator SNR (b) Output CIC filter SNR

The CIC filter Verilog code was written and simulated by Matlab software. The signal to noise ratio is 141.56 dB in sigma delta modulator output and it is increased to 145.35 dB in the decimation stages. To improve the signal to noise ratio, word length of recursive CIC filter should be increased but the speed of filter calculation is also decreased.

The chip layout on Virtex II FPGA board has been shown in Figure 9.

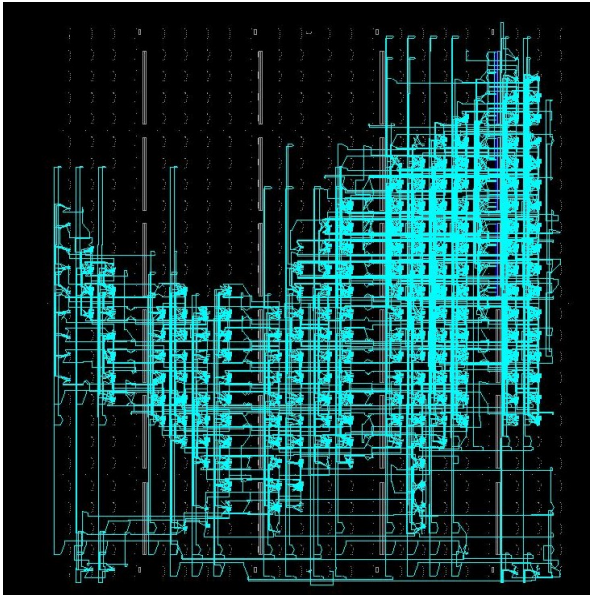

Fig. 9 The core layout on FPGA board

## IV. CONCLUSION

Recursive CIC filters have been designed and investigated. Enhanced high Speed CIC filters was obtained by three ways. The pipeline structure, using the modified carry look-ahead adder (MCLA) and truncation lead us to have high speed CIC filter with the maximum throughput of 190 MHz. The evaluation indicates that the pipelined CIC filter with MCLA adder is attractive due to high speed when both the decimation ratio and filter order are not high as stated in the Hogenauer Comb filter. Since the first stage of the CIC filter require maximum word length and also because of the recursive loop in its structure, the reduction in power consumption is limited by the throughput. Thus the truncation will reduce the power consumption and the number of calculation. The power consumption computed using CAD tools (Cadence and Synopsys) and 0.18 μm Silterra technology library gives 3.5 mW power consumption at maximum clock frequency.